\documentstyle[emulateapj]{article}

\def\lsim{\ \raise -2.truept\hbox{\rlap{\hbox{$\sim$}}\raise5.truept  
\hbox{$<$}\ }}                                                          
\def\gsim{\ \raise -2.truept\hbox{\rlap{\hbox{$\sim$}}\raise5.truept  
\hbox{$>$}\ }}                                                          
\def\msun{M_{\odot}}

\begin{document}

\submitted{Accepted for publication in ApJ Letters}

\lefthead{CAVALIERE \& MALQUORI} 
\righthead{EVOLUTION OF BL LAC OBJECTS}

\title{The Evolution of the BL Lac Objects}

\author{A. Cavaliere$^1$ and D. Malquori$^{2,3}$}

\altaffiltext{1}{ Astrofisica, Dipartimento di Fisica, 2a Universit\`a 
degli studi di Roma, Roma, 00133, Italy}
\altaffiltext{2}{Dipartimento di Astronomia, Universit\`a di Padova, 
Padova, 35122, Italy}
\altaffiltext{3}{Institut d'Astrophysique de Paris (CNRS),  
Paris, 75014, France}

\begin{abstract}

The BL Lac population is set apart from the rest of the active galactic
nuclei by a number of peculiar features, including remarkably little 
signs of cosmological evolution.
We focus on this feature, and take hint from the observations 
to reconsider extraction of the primary energy from a Kerr hole 
via the Blandford-Znajek (BZ) process; in fact, we stress how this can power 
the emitting jets of the BL Lac objects in a regime of low accretion.
We show that the timescale for the BZ process to exhaust the  
rotational energy of the hole spans many gigayears in all such sources. 
We compute scales long and uniform enough to yield an intrinsically slow 
population evolution, giving rise to number counts of ``Euclidean'' or 
flatter shape; even longer timescales apply to the additional, 
analogous process involving the inner accretion disk directly.
Such slow evolution takes place in ambients and at epochs ($z \lsim 1$) 
where interactions of the host galaxies with companions become rare or weak.
On the other hand, at higher $z$, we argue that frequent interactions of the 
hosts within groups enhanced, each for some 
$10^{-1}$ Gyr, 
the accretion rates and the emissions from the disks, attaining power levels 
that are typical of the radio-loud quasars.
But such interactions were also likely to disorder the large-scale 
magnetic field in the disk and to decrease the angular momentum of the hole, 
thus limiting 
the lifetime of all BZ emissions to under a few gigayears. 
Then the stronger evolution of the flat-spectrum radio-loud quasars 
was driven, similar to the radio-quiet ones, 
by the interaction rate decreasing rapidly with $z$. 

\end{abstract}


\section{Introduction}

The BL Lacertae objects among the active galactic nuclei (AGNs) exhibit a 
number of peculiarities 
(see Urry \& Padovani 1995, Sambruna 1999, and references therein), namely: 
$(a)$ lack of emission lines with equivalent width exceeding 5 \AA; 
$(b)$ a pure nonthermal continuum; 
$(c)$ flat-spectrum radio emission, and powerful $\gamma$-rays that extend 
up to TeV energies in some objects; 
$(d)$ rapid variability; 
and $(e)$ strong  and highly variable optical polarization. 
To some extent, they share the three latter features with the 
flat-spectrum radio-loud quasars (FSRQs).

The current understanding links the peculiarities $c,d$, and $e$ to 
synchrotron and inverse Compton emissions from a relativistic jet 
with bulk Lorentz factors $\Gamma \sim 3-10$, viewed at a small angle 
of order $\Gamma^{-1}$ (see Ghisellini et al. 1993). 
On the other hand, the peculiarities $a$ and $b$ are widely related to a 
dearth of gas surrounding the activity center at both parsec and 
10 $\mu$pc scales, and/or to the lack of an adequate nonbeamed continuum.
An additional peculiarity is constituted by the behavior of the population,   
which shows  {\it little or no} signs of evolution 
(see Stickel et al. 1991; Wolter et al. 1991; Bade et al. 1998), 
at striking variance with the rest of the AGNs. 

To us, the peculiarities $a$ and $b$ suggest that current accretion 
is {\it low} in BL Lac objects, providing little gas and a weak ionizing 
continuum. 
Taking up previous proposals (see Blandford 1993 and references therein), 
we consider the main power source for the BL Lac emissions 
to be provided by the extraction in electromagnetic form of 
{\it rotational} energy, stockpiled in the central black hole 
and associated with the inner accretion disk.  
We develop this view to explain the other BL Lac peculiarities,  
and specifically the weak evolution. 

\section{Evolutionary behavior}

Once Doppler enhancements of order $ \Gamma^{3-4}$ are accounted for, 
the observations of BL Lac objects (see Sambruna, Maraschi \& Urry 1996), 
indicate typical rest frame luminosities 
$L \lsim 10^{44}$ ergs s$^{-1}$, 
which are moderate on the quasar standards; 
meanwhile, the rotational energies extractable 
from Kerr holes (KHs) are around $E \sim 5 \, 10^{61}M_8$ ergs. 
So their exhaustion would take in individual objects times 
$\tau \sim E/L$ of many gigayears, which are quite {\it longer} 
than the flashes lasting about $ 10^{-1}$ Gyr of standard quasars 
fueled by accretion (see Cavaliere \& Vittorini 1998). 
In the few instances of larger $L$, the Doppler factors inferred 
from the data in the framework of the synchrotron self-Compton 
radiation -- which constitute lower bounds anyway -- are 
particularly low and uncertain pending confirmation from 
superluminal velocities (P. Padovani 1998, private communication).
The long debated kinetic component of the jets is unlikely to increase 
the power budget of the typical BL Lac objects beyond some 
$10^{45}$ ergs s$^{-1}$ (Ghisellini \& Celotti 1998). 

The population-averaged $\langle L(t) \rangle$ is constituted by 
the convolution of the individual behaviors. 
When these are all similar, as we shall show below, 
$\langle L(t) \rangle$ also shares the same timescale $\tau$.
Heuristically, a long $\tau$ will imply little population evolution 
and, therefore, relatively flat number counts,  
or equivalently $V/V_{max}$ close to $0.5$. 

A quantitative statement can be made in simple terms on recalling that 
the integral counts $N(>S)$ at high fluxes S may be cast into the form 
(see Weinberg 1972): 
\begin{equation}
 N(>S)\propto S^{-3/2}[1-C(S_o/S)^{1/2}+0(S^{-1})] ~~.
\end{equation}
In turn, C can be expressed (Cavaliere \& Maccacaro 1990) in terms of the time 
scale ${\tau_e}$ for population changes by
\begin{equation}
C= \frac{3\, D_o <l^2>}
{4\, R_H<l^{3/2}>} \,\; [2(1+\alpha)-\frac{1}{H_o \tau_e}] ~~. 
\end{equation}
Here  $R_H = 3000\, h^{-1}$ Mpc is the Hubble radius with the Hubble 
constant $h = H_o /100$ km s$^{-1}$ Mpc$^{-1}$, 
$D_o\equiv (P_o/{4 \pi S_o)^{1/2}} \approx 200\, h^{-1}$  Mpc 
in terms of the power and flux of the brightest BL Lac objects, 
$<l^n>$ is the (normalized) $n$-th moment of the luminosity function (LF), 
and $\alpha$ is the spectral index.

\vspace{9.2cm} 
\includegraphics{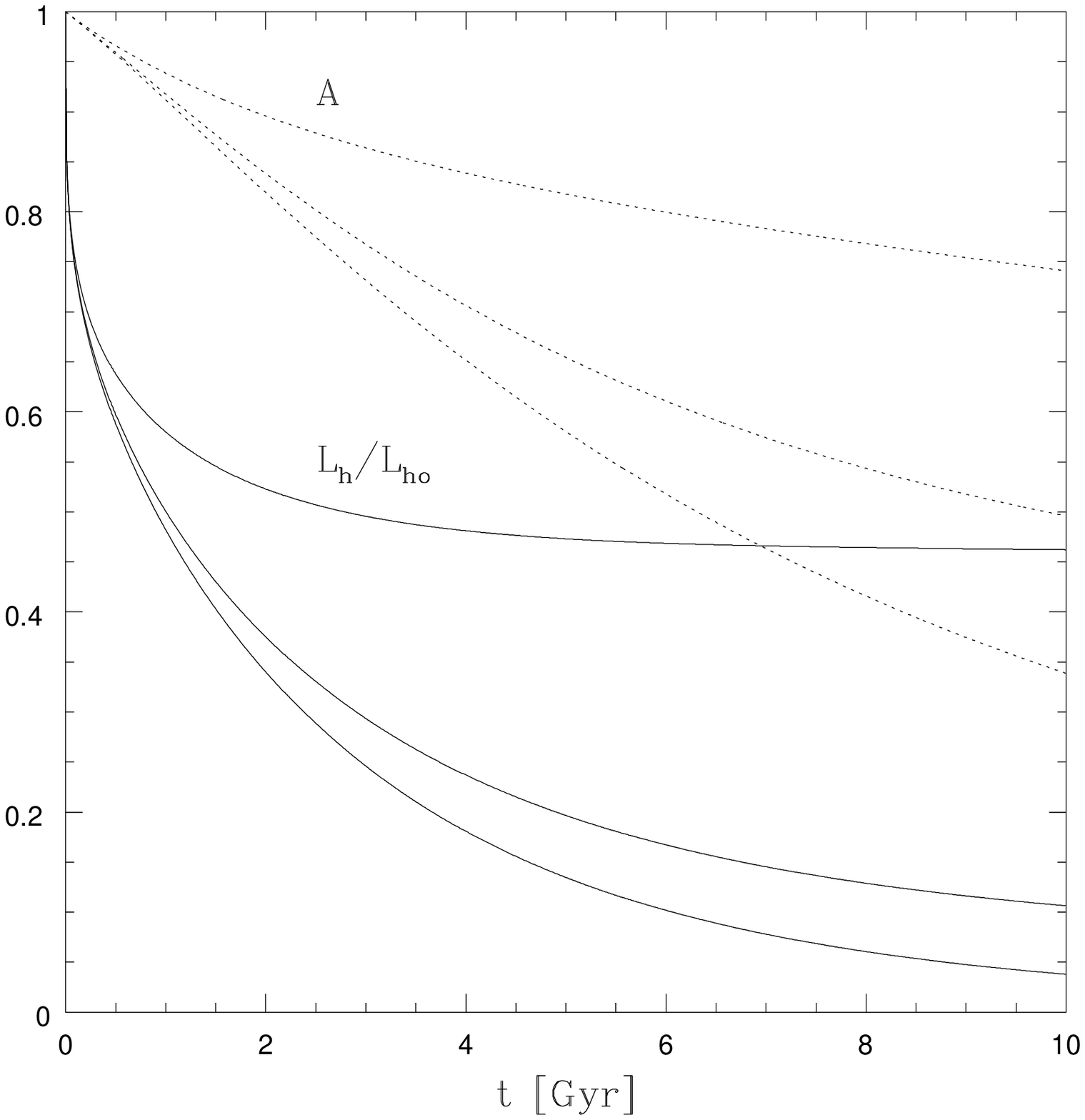} 
{\footnotesize FIG. 1 -- The runs of $A(t)$ and of $L(t)$ for 
$\dot M=$ const, corresponding to initial normalized accretion rates 
$\dot m_o =5 \, 10^{-2}$ (upper curves), $10^{-2}$ (middle curves), 
and $10^{-3}$ (lower curves). 
Note the convergent behaviors at the lowest rates. }
\vspace{\baselineskip}

The evolutionary rate 
$\tau_e^{-1} = \tau_{D}^{-1}+ (\beta-1) \tau_L^{-1}$ 
comprises two contributions that are due to density and to luminosity 
evolutions, respectively; the latter is amplified by a steep LF (modeled
with $L^{-\beta}$ in the relevant range), and $\tau_D$ 
is negative in case of negative evolution. 
 From eq. (2) it is seen that the bright counts tend to the 
``Euclidean'' shape $N(>S) \propto S^{-3/2}$ or flatter even when 
$\beta \gsim 2.5$, and even in the presence of positive evolution, 
provided this is slow enough to statisfy 
$\tau_e\, H_o \geq [2(1+\alpha)]^{-1}$. 

\section{A simple model: beams from Kerr holes} 

We begin with considering the specific model where the primary power 
is extracted from the rotational energy of a KH, 
following Blandford \& Znajek (1977).
They tackled a stationary, axisymmetric and force-free 
magnetosphere surrounding the KH, and derived the scaling 
$L_h \propto  B^2 \, r_h^2\, A^2$ for the power extracted from the KH 
{\it itself} in the form of Poynting flux; here  $A\leq 1 $ denotes 
the angular momentum of the hole, $J$, in units of $G M^2/c$, 
and $B$ is the magnetic field threading the horizon at  
$r_h=[1+(1-A^2))^{1/2}] \, G M/c^2$.  

The strength of the field that is sustainable at the horizon constitutes 
a delicate issue.
A recent reappraisal by Ghosh \& Abramowicz (1997) has been based 
on continuity of the magnetic pressure at the inner edge of the 
accretion disk and on numerical computations -- which are self-consistent, 
if with  dynamic range limited to small and medium scales -- 
of its MHD and thermal structure. 
They find that $B^2/{8 \pi} \sim \alpha p $ holds for standard 
$\alpha$-disks in terms of the maximal pressure at the inner rim; 
they (conservatively) evaluate the resulting power to read 
\begin{equation}
L_h \simeq ~ 
\begin{array}{l}
2\, 10^{44} M_8\, A^2  ~ ergs ~ s^{-1} \\
8 \; 10^{42} {M_8}^{11/10} {{\dot m}_{-4}}^{4/5} A^2 ~ ergs ~ s^{-1} ~,
\end{array} 
\end{equation}
when the inner region is dominated, respectively, by radiation 
[$p\propto (\alpha \,M)^{-1}$] or by gas pressure 
[$p\propto (\alpha\, M)^{-9/10} {\dot m}^{4/5}$].
We recall from Novikov \& Thorne (1973) that only when 
the accretion rate in Eddington units exceeds 
${\dot m} \equiv {\dot M} c^2/ L_E \simeq 3\, 10^{-4} \, M^{-1/8}_8$ 
(for $A =1$) does the transitional radius $r_* \propto {\dot M}^{16/21}$  
(separating the radiation-dominated region from the gas-dominated region) 
exceed the radius $r_{ms}$ of the marginally stable orbit so that an inner, 
radiation pressure dominated region in fact occurs. 
Then the ratio of the Blandford-Znajek (BZ) emission to the 
thermal luminosity scales as 
$L_h/L_{th} \approx 3\, 10^{-2}\, A^2/ {\dot m}$. 
Thus, in the range $10^{-3} \lsim \dot m \lsim 10^{-2}$, the rest 
frame power $L_h$ is adequate to sustain the nonthermal, beamed emissions 
of BL Lac objects and to match or dominate the thermal component $L_{th}$.

As for the evolutionary behavior of the population, we first consider the 
time dependence of  $L_h(t)$ for an individual source; this is related, 
by equations (3), to the behaviors of $M(t)$ and of $J(t)$. 
The key point is that accretion of $J$ also takes place along with $\dot M$, 
and is always important, as first pointed out by Bardeen (1970), 
being related to $\dot M$ by the conditions in the disk at $r_{ms}$. 

In fact, the run of $M(t)$ is described by
\begin{equation}
{dM \over dt} = e_{ms} \dot M - L_h/c^2 ~, 
\end{equation}
where ${e}_{ms} \simeq 1$ is the adimensional specific energy 
at $r_{ms}$; actually, ${e}_{ms} = 0.58$ holds for $A=1$, 
rising to $0.94$ for $A = 0$ (de Felice 1968; Bardeen 1970). 
On the other hand, the run of $J(t)$ is coupled to the former by 
\begin{equation}
{dJ \over dt} =  j_{ms}  \dot M  - {L_h \over {k\, \Omega_h}} ~  
\end{equation}
(Malquori 1996; Moderski \& Sikora 1996). 
Here $j_{ms}$ is the specific angular momentum at $r_{ms}$, with values 
close to $ 2\,  e_{ms}\, c\, r_h$; moreover, $\Omega_h \equiv A c/2 r_h$, 
while $k = 1/2$ for maximum power extraction. 

On the right-hand side of equation (5) the accretion of $J$ can be balanced 
out by the loss due to the BZ process, causing $dJ/dt$ to vanish for values 
of $A$ around $ 50 \,\dot m$,  given $10^{-3} \lsim \dot m \lsim 10^{-2}$. 
Instead, $A(t) \propto J(t)/M^2(t)$ decreases monotonically (see Fig. 1).
Correspondingly, $L_h(t)$ (shown in Fig. 1 with all relativistic factors
and detailed $A$-dependencies in place) decreases steeply at first 
because of the relativistic factors, but then it tends to level off 
(or to increase again for high  $\dot M$) because of the hole mass increase; 
in between, the decrease takes place on the scale 
${\tau} \equiv L/\dot L \simeq   J/2 \dot J$.  
Such behavior and scale are {\it similar} for all objects at low 
$\dot m$ where the BZ power dominates. 

In sum, the sources powered by KHs fade slowly, but remain above levels 
$L_{ho}/10  \sim  10^{44}$ erg s$^{-1}$ that are relevant for 
BL Lac objects for times $\tau \gsim 8$ Gyr. 
So their population undergoes a slow luminosity evolution; the condition 
$\tau_e\, H_o  \simeq [2(1+\alpha)]^{-1}$ tends to hold especially for 
$h \simeq 0.7$, and the bright counts in {\it any} band ought to be quite 
{\it flatter} than with the quasars. 

As an example, we compute in detail eqs. (1) and (2) for the radio band 
on the basis of the above luminosity evolution alone.
In the radio, the BL Lac counts are arguably least sensitive to the spectral 
changes with power proposed by Fossati et al. (1998; 
see also Georgenopoulos \& Marscher 1998 and Rector et al. 1999).  
There $\alpha \simeq 0$ and $ \beta \simeq 2.5$ apply 
(we refer to the model 
LF computed by Padovani \& Giommi 1995), and the condition 
$\tau_e\, H_o  \simeq 1/2$ for nearly Euclidean counts is closely met. 
In fact, we find the shape shown in Figure 2 by the solid line; 
this is easily consistent with the best fit ({\it dashed line}) to the counts 
by Stickel et. al. (1991), a published sample that is complete 
although it is affected by much statistical variance. 
A flatter LF will yield counts flatter still. 
On the other hand, 
a flatter LF, but with fast evolution, would yield steep counts 
as illustrated by the dotted line in Figure 2. 

Actually, our results provide an upper limit to the counts,  
since  the values of $\tau$ evaluated above provide minimal 
timescales while an additional negative density evolution is likely to 
occur, as discussed next. 
More extensive and deeper data will be released soon 
(see Giommi, Menna \& Padovani 1999 and Rector et al. 1999)  
and will provide more challenging evolution indicators for our prediction. 

\section{Discussion and conclusions} 

{\it All} peculiarities of the BL Lac objects are consistent with 
the extraction in {\it electromagnetic} form of the {\it rotational} 
energy stockpiled in a KH, in a regime of {\it low} accretion. 
In fact, a primary Poynting flux, as produced via the BZ process by 
the KH itself, is attractive (Ghisellini \& Celotti 1998) because it is 
energetically adequate and is conducive to jets with bulk $\Gamma$ $ \sim 10$; 
by way of the implied large-scale electric fields, it is conducive also 
to distributed and continuous acceleration up to isotropic electron 
energies in the range of $10^2$ GeV. 
Thus, highly beamed, nonthermal (and polarized) continua extending 
into the TeV $\gamma$-ray range can be generated; 
if a jet is oriented close to our line of sight, high apparent power and 
fast time variability will be observed (Begelman et al. 1984). 
In turn, {\it low} current accretion rates, $\dot m \lsim 10^{- 2}$, 
are consistent with negligible dilution by thermal continua 
(then also unable to degrade the TeV $\gamma$-rays) 
and with emission lines of small equivalent width. 
But our thrust has been to show how this view leads us to predict 
intrinsically {\it slow} population evolution.

Actually, an additional MHD emission $L_d$ is produced by the inner disk 
(Blandford \& Znajek 1977; Livio, Ogilvie \& Pringle 1999).
This scales still with $B^2$, but also with $r^{3/2}_{*} r^{1/2}_h$ 
so that the ratio $L_d/L_h \propto (r_{*}/ r_h)^{3/2} 
\propto \dot M^{1.15}$; the coefficient is independent of 
$A^2(t)$ and somewhat larger than in eq. (3a). 
Consideration of $L_d$ does not affect materially our evaluations as long as 
$\dot m \lsim 10^{-2}$, and so $L_d \sim L_h$, which is appropriate to the 
BL Lac objects.  
Then the KH and the disk are dynamically and magnetically linked 
nearly in one system in which -- within (uncertain) factors of a few -- 
the hole still may be regarded as the energetic flywheel.

\vspace{9.2cm} 
\includegraphics{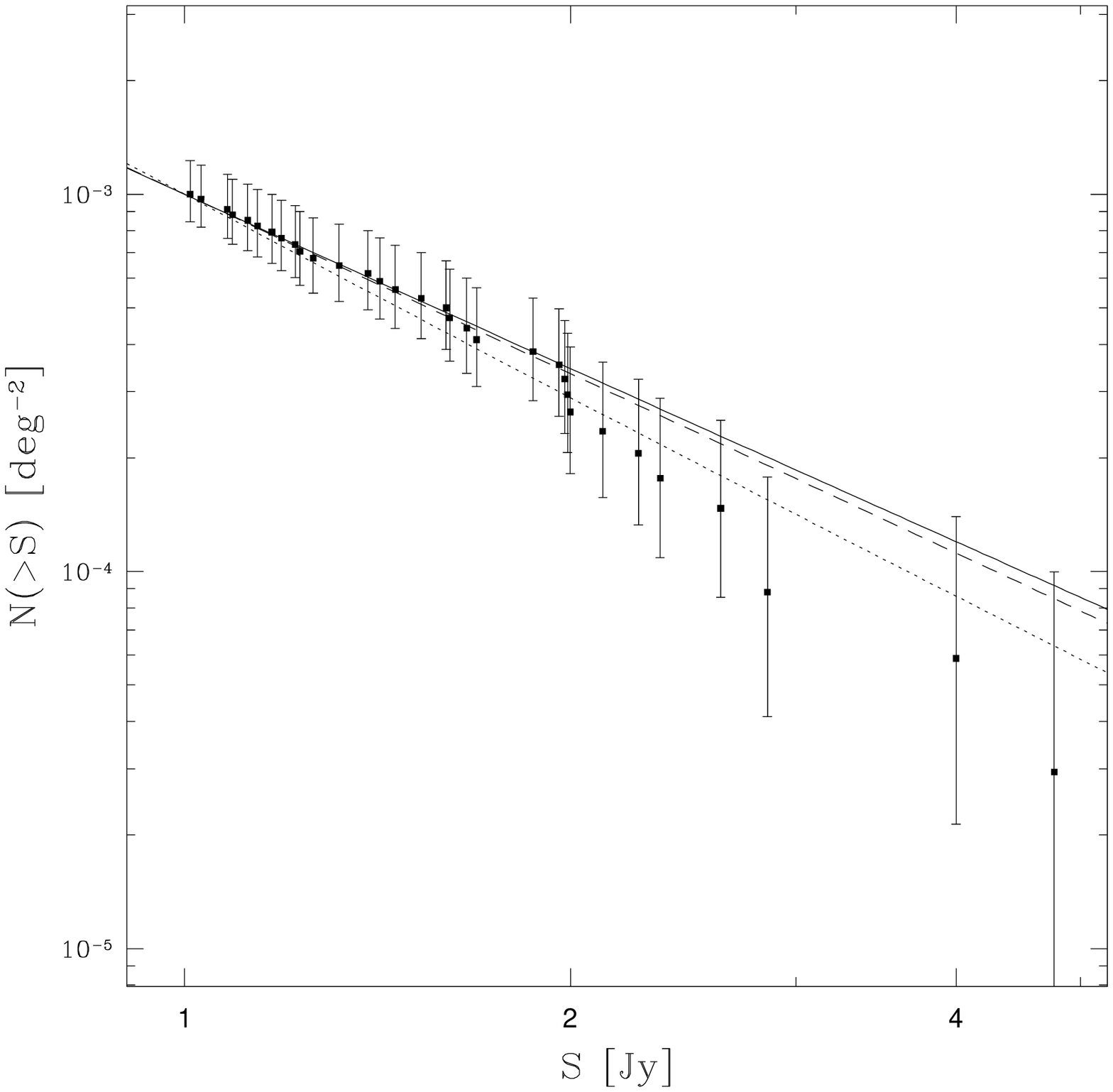} 
{\footnotesize FIG. 2 -- BL Lac counts at 5 GHz, from Stickel et al. 1991.
The dashed line reproduces their best fit $N(>S) \propto S^{-1.6}$
(the actual slope is $1.58 \pm 0.44$ at a $1\, \sigma$ level). 
The solid line gives our result for pure luminosity evolution with 
$\tau = 8$ Gyr, assuming constant $L_{radio}/L_h$ and $\beta= 2.5$. 
The dotted line represents for comparison a case where 
$\beta= 2.2$ and $\tau = 2$ Gyr.}
\vspace{\baselineskip}
 
The stand in this context of the other Blazars, the FSRQs, 
is interesting as such objects exhibit features similar to the BL Lacs, 
less pure but with larger outputs. 
Tapping the rotational energy in order to produce relativistic jets, 
blazing beams, and nonthermal luminosities is tantalizing also for these 
objects, but there are also important differences. 

Now the maximal KH power $L_h$ from eq. (3a) is inadequate -- 
with masses limited to $M \lsim 4 10^9 \, h^{-1}\, \msun$ 
(see Magorrian et al. 1998) -- to match powers of 
$ 10^{46}\, h^{-2}$ erg s$^{-1}$ or more;  
a larger disk component $L_d$ to the BZ emission is cogently called for. 
In fact, $L_d \gg L_h$ now holds (with the KH now slaved to the disk)
since $\dot m \sim 1$; we know this is to hold for the FSRQs, because their 
emission lines and thermal components, which dilute the BL Lac-like 
peculiarities to approach the standard quasar activity, 
do signal currently {\it high} accretion rates. 
We also expect the larger disk component in FSRQs to imply -- 
relative to the BL Lac conditions -- more plasma lifted up  
and electric fields more accurately screened out; then 
the electrons not only may loose their energy faster at such larger total 
outputs (Fossati et al. 1998), but are also accelerated to lower energies, 
which is another, intrinsic reason to emit softer $\gamma$-rays. 

Since the BZ emissions are marked by long timescales at constant $\dot M$, 
the long persistence of radio-emitting, 
rotationally driven jets might be seen as statistically embarassing 
(see Moderski, Sikora \& Lasota 1998). 
However, we argue that the actual timescales of the FSRQs are much shorter 
because of the boundary conditions set to such high values of 
$\dot M$ by the host galaxy. 
For the sake of this argument, we shall discuss first the scenario (most 
constraining to us) where the BL Lac objects are descendants of FSRQs.

We recall from Cavaliere \& Vittorini (1998) that {\it recurrent}
interactions and merging events of the E host galaxies with  satellites or 
with group companions (for which there is plenty of morphological and 
kinematical evidence; see Bertola \& Corsini 1998) 
are necessary to cause substantial losses of angular momentum,  
$|\Delta j|/j \sim 1$, of the gas residing in either interaction partner;  
such events every $1-2 $ Gyr trigger for 
some $10^{-1}$ Gyr gas inflow toward the nucleus, which is conducive to 
$\dot M$ and also to accretion of angular momentum onto the hole. 
We concur with Moderski, Sikora \& Lasota (1998) that the {\it sign} 
of such imported momentum relative to the previous one makes a difference
as for disk structure and for the KH rotation. 
We stress that each strong interaction has a $1/2$ chance to accrete 
a component opposite  to the initial {\bf J} of the KH and of the disk, 
thus wrecking the latter's large-scale magnetic field and spinning the 
former down, effectively quenching  their emissions. 
When the disk is rebuilt, the internal conditions for large-scale 
{\bf B} and those in the host galaxy for beam propagation 
(see Livio 1999) may not be met again. 

Thus we hold that the probability is small for combining persistent extraction 
of BZ power with frequent episodes of high $\dot m$. 
But the quasar evolution since $z \approx 3$ is governed just by 
high-$\dot m$ episodes driven by {\it interactions}; these are  
initially frequent and strong in the era of group formation, but then 
drop off in rate and strength with decreasing $z$ as groups are 
reshuffled into clusters (Cavaliere \& Vittorini 1998). 
As long as such interactions are effective, 
they will intermittently switch off all BZ outputs from the KH and the disk. 

We conclude that strong evolution and long-lived extraction of 
BZ power entertain a {\it complementarity} relationship. 
Indeed, the rotational power and the beams it feeds  persist 
safely for many gigayears only in a regime of permanently low 
accretion, which is what prevails for 
$z \lsim 1.5$, after a ``last interaction'' setting  $A \simeq 1$; 
such conditions will lead to the BL Lac phenomenon. 
At higher $z$, the most likely activity pattern
was that corresponding to FSRQs, with dominant
thermal and disk BZ emissions limited to a few gigayears by 
initially frequent, but then subsiding, interactions; 
the evolution of such sources is then expected to be close (cf. Goldschmidt 
et al. 1999) to the standard quasars', being driven by the same 
$z$-dependent $\langle \dot m \rangle$ only somewhat slowed down by the 
beam attempts to persist in between interactions.
On the other hand, if it is maintained (see Jackson \& Wall 1999) that 
BL Lacs and FSRQs belong to different and secluded lineages, 
then little  constraints  will
arise from FSRQs for the the present view of the  
astrophysics and evolution of BL Lac objects.    

Finally, enlarging our view of this recently rejuvenated field above 
details and beyond the theoretical dream of a relativistic homopolar 
inductor, we summarize and emphasize the two key issues in this context. 
First and concerning the BL Lac population, up to now  
flat counts and $V/V_{max}\simeq 0.5$ persistently signal a {\it {weaker}} 
evolution than with the quasars. 
This indicates either a slower luminosity evolution with 
timescales $\tau_L/(\beta -1) $ of several gigayears, as considered above, 
or a negative density evolution with $\tau_D < 0$, or some mix of the two 
(some negative density evolution is also implied  from BL Lac 
birth since $z \approx 1.5$).
Second, concerning the source structure, the current wisdom 
on disks (but ADAF or ADIOS flows, possibly setting in at low $\dot m$, as 
discussed by Blandford \& Begelman 1998, have yet to be firmly fit in)
has that the disk radiation pressure is independent of accretion for 
$\dot m$ above some $10^{-3}$, since then $p \propto T^4$ holds 
in the inner region. {\it As long as} good coupling of radiation, particles 
and large-scale magnetic fields can be retained, this leads to a {\it large} 
ratio $(L_h + L_d)/ L_{th} \propto 1/\dot m$;   
in addition, as $\dot m$ approches $10^{-3}$ from above, $r_*$ shrinks to 
approach $r_{ms}$, and {\it comparable} $L_h \sim L_d$ obtain.  
Such conditions are best suited to yield pure nonthermal radiation 
of power moderate but extending up to 
TeV $\gamma$-rays, 
and by the same token they yield persistent beams with intrinsically weak 
evolution, precisely the features marking out the BL Lac objects. 

\section*{Acknowledgments}
\noindent
We thank V. D'Elia and P. Padovani for several useful discussions, 
and our referee for helpful comments. 
A.C's share of the work was completed while he visited the 
Asiago Observatories. 
The grants were from ASI and MURST.

E\_mail: cavaliere@roma2.infn.it; malquori@pd.astro.it

\end{document}